# Self-repairing high entropy oxides

Zongwen Liu[1,2]*†, Pengru Huang[3,4]†, Lixian Sun[4]*, Yanping Liu[5], Jiangtao Qu[6], Julie Cairney[6,7], Zhong Zheng[1], Zhiming M. Wang[8], Naveed A. Khan[1], Zhiping Lai[9], Li Fu[10], Bing Teng[11], Cuifeng Zhou[1], Hong Zhao[1], Fen Xu[4], Pan Xiong[12], Junwu Zhu[12], Peng Yuan[13], Kosta Tsoutas[14], Behnam Akhavan[14,15], Marcela M. Bilek[2,14,15], Simon P. Ringer[6], Kostya S. Novoselov[3]*

[1]School of Chemical and Biomolecular Engineering, The University of Sydney; NSW 2006, Australia.

[2]The University of Sydney Nano Institute, The University of Sydney; NSW 2006 Australia.

[3]Department of Materials Science and Engineering, National University of Singapore; 117575, Singapore.

[4]Guangxi Key Laboratory of Information Materials, Guangxi Collaborative Innovation Center of Structure and Property for New Energy and Materials, School of Material Science & Engineering, Guilin University of Electronic Technology, Guilin, 541004, PR China.

[5]School of Physics and Electronics, Hunan Key Laboratory for Super-microstructure and Ultrafast Process, Central South University; Hunan 410083, China.

[6]School of Aerospace, Mechanical and Mechatronic Engineering, The University of Sydney; Sydney, NSW 2006, Australia.

[7]The Australian Centre for Microscopy and Microanalysis, The University of Sydney; NSW 2006 Australia.

[8]Institute of Fundamental and Frontier Sciences, University of Electronic Science and Technology of China; Chengdu 610054, China.

[9]Advanced Membranes and Porous Materials Center, Division of Physical Science and Engineering, King Abdullah University of Science and Technology; Thuwal 23955, Saudi Arabia.

[10]State Key Laboratory of Solidification Processing, Northwestern Polytechnical University; Xi'an 710072, China.

[11]College of Physics, Qingdao University; Qingdao 266071, China.

[12]Key Laboratory for Soft Chemistry and Functional Materials of Ministry Education, Nanjing University of Science and Technology; Nanjing 210094, China.

[13]CAS Key Laboratory of Mineralogy and Metallogeny, Guangdong Provincial Key laboratory of Mineral Physics and Materials, Guangzhou Institute of Geochemistry, Chinese Academy of Sciences; Guangzhou 510640, China.

[14]School of Physics, The University of Sydney; NSW 2006, Australia.

[15]School of Biomedical Engineering, The University of Sydney; NSW 2006, Australia.

*Corresponding authors. Email: zongwen.liu@sydney.edu.au; sunlx@guet.edu.cn; kostya@nus.edu.sg

†These authors contributed equally to this work.

**Abstract:** All biological organisms, from plants to living creatures, can heal minor wounds and damage. The realization of a similar self-healing capacity in inorganic materials has been a design target for many decades. This would represent a breakthrough in materials engineering, enabling many novel technological applications, since such materials would be able to resist damage caused by electromagnetic irradiation and/or mechanical impact. Here we demonstrate that a high-entropy oxide is intrinsically capable of undergoing an autonomous self-repairing process. Transmission electron microscopy revealed that the spinel structure of $(AlCoCrCu_{0.5}FeNi)_3O_4$ can regrow and repair itself at the atomic level when damaged. Density functional theory calculations reveal that the extra enthalpy stored in the high entropy material during fabrication can be released to effectively heal macroscopic defects by regrowing into a partially ordered state. This extraordinary self-repairing phenomenon makes this new material highly desirable as a coating, enabling structures used in harsh environments to better withstand damage, such as cosmic irradiation in space, nuclear irradiation in nuclear power facilities, or tribological damage. Most importantly, our results set the general design principles for the synthesis of self-repairing materials.

**One-Sentence Summary:** Experimental results and theoretical calculations are presented to establish principles for the design of a new class of high entropy oxide materials with an extraordinary capacity for self-repair following damage caused by electromagnetic irradiation or mechanical impact.

The development of high entropy alloys (HEAs) nearly two decades ago (*1, 2*) opened a new area of materials design through control of the configurational entropy (*3, 4*). HEAs exhibit high strength (*5, 6*), high thermal stability and exceptional anti-corrosion properties (*7*). Due to high atomic-level stresses and chemical heterogeneity, they are also highly irradiation resistant (*8-11*). Recently, there have been increasing studies involving high entropy materials of other systems (*12, 13*) with various forms of other high entropy materials being created, including ultrahigh-temperature borides (*14*), chalcogenides (*15*), carbides (*16*) sulfides (*17*), nitrides (*18, 19*), fluorite oxides (*20, 21*), perovskite oxides (*22, 23*) and spinel oxides (*24-27*).

However, there are currently no detailed studies to date that investigate the influence of the balance between enthalpy and entropy on the structural properties of such materials. During material synthesis, which usually occur at high temperatures, entropy predominates, resulting in a mixed state. On the other hand, at lower temperatures the influence of the configurational entropy diminishes, resulting in the release of the enthalpy stored in the oxide (due to chemical interactions between the adjacent atoms which is of a different nature when in the disordered state) and the simultaneous reconfiguration to a more ordered state. Potentially this energy stored could be used for defect reconstruction or self-repairing.

In this work while we focus on the self-repairing behavior of a high entropy oxide (HEO) based on the six common elements of Al, Co, Cr, Cu, Fe and Ni (*26*), we believe that our results are generic for other systems of high entropy materials. HEAs containing these six metallic elements were first reported by Tung et al. in the bulk form having been prepared by the arc melting and casting methods (*6*). Recently, a low Cu variant AlCoCrCu$_{0.5}$FeNi HEA was fabricated by radio frequency (RF) magnetron sputtering in the form of thin films (*28, 29*). Huang et al (*30*) performed the first attempt to introduce oxygen into the AlCoCrCu$_{0.5}$FeNi system. This represents one of the earliest works to have fabricated HEO thin films containing multiple metallic elements. Later, Chen and Wong synthesized the Al$_x$CoCrCuFeNi oxide films with a spinel structure by reactive sputtering (*31, 32*).

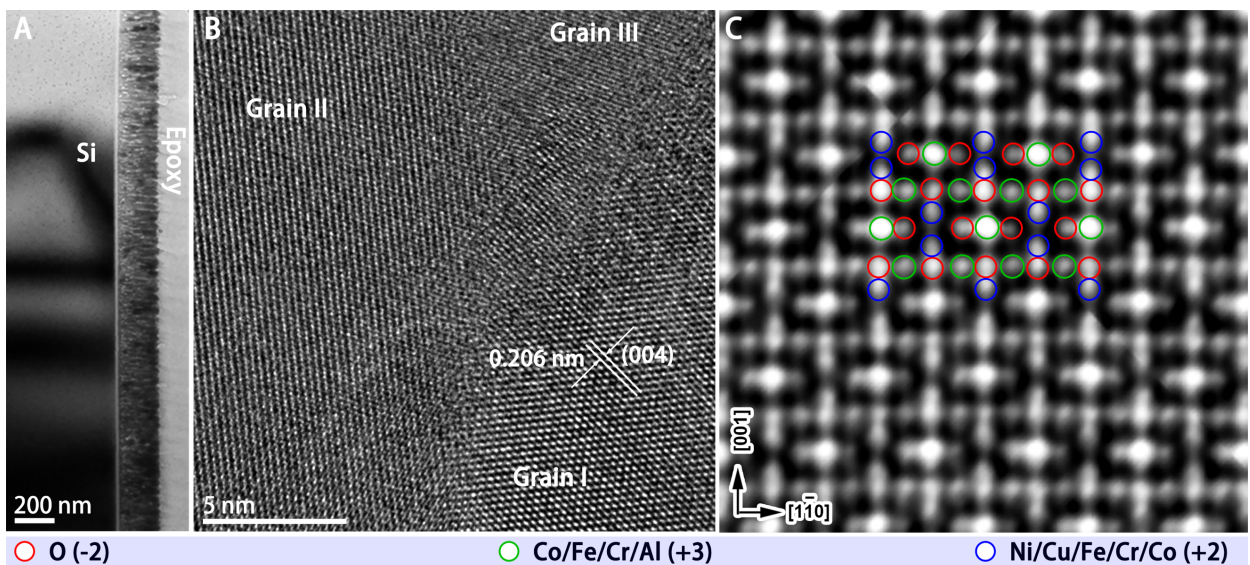

**Fig. 1. TEM micrographs and structural characterization of the (AlCoCrCu$_{0.5}$FeNi)$_3$O$_4$ high entropy oxide thin film**. (**A**) Bright-field cross-sectional image showing the thin film with a thickness of about 200 nm viewed at edge-on orientation of Si [011]. (**B**) High-resolution TEM image of the oxide thin film showing three adjacent nanocrystals. (**C**) iDPC image of a nanocrystal showing the spinel projection along [110] orientation with the oxygen atom columns being resolved clearly. The red circles indicate columns of oxygen atoms and green circles indicate columns of trivalent atoms, while the blue circles indicate columns of divalent atoms.

To demonstrate the self-repairing behavior, we synthesized the HEO of (AlCoCrCu$_{0.5}$FeNi)$_3$O$_4$, which possesses a spinel structure (*27*). The (AlCoCrCu$_{0.5}$FeNi)$_3$O$_4$ film was deposited on a Si substrate by RF magnetron sputtering from a single stoichiometric target in a mixed Ar and O$_2$ atmosphere. The film grew to a thickness of ~200 nm in ~2 hours under the deposition power of 150W (Fig. 1A) and is composed of randomly oriented nanocrystals ranging from 10-40 nm in diameter (Fig 1B). Our recent study (*27*) involving electron diffraction and X-ray diffraction analysis demonstrated that these nanocrystals possess a spinel cubic structure with a lattice parameter of $a$ = 0.823 nm. Fig. 1C is an integrated differential phase contrast (iDPC) scanning transmission electron microscopy (STEM) image revealing the spinel structure in the [110] orientation. The oxygen atom columns were clearly resolved (red circles) in this iDPC image, recorded using a monochromated double aberration corrected TEM (FEI Themis Z300).

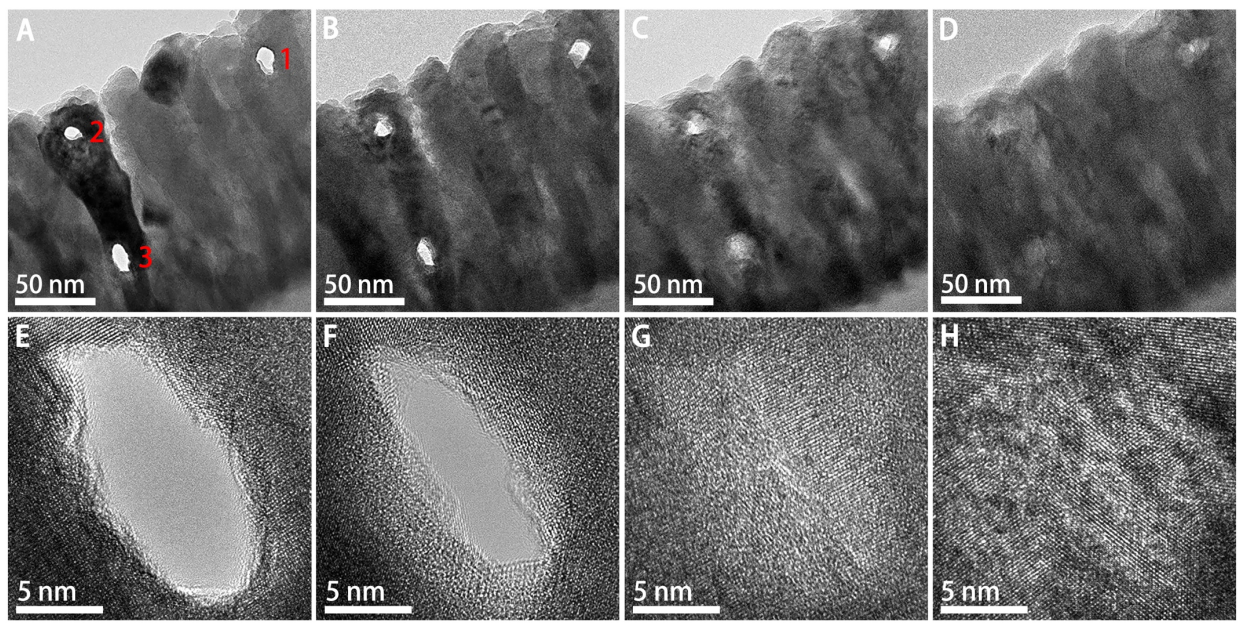

**Fig. 2. The self-repairing of the AlCoCrCu$_{0.5}$FeNi)$_3$O$_4$ oxide film in air at room temperature**. (**A**) A low mag image showing three holes, marked 1, 2 and 3 with diameters of 15-20 nm, drilled with a converged 300kV electron beam. (**B**) An image showing the 3 holes one week after drilling. (**C**) An image showing the 3 holes two weeks after drilling. It is very obvious to see the filling of the holes via a self-repairing reaction. (**D**) Three weeks after the holes were drilled, all 3 holes have largely closed. (**E**) A HRTEM image of hole 3 immediately after it had been 'drilled'. (**F**) After one week, hole 3 was reduced to ~half its original diameter. (**G**) Hole 3 was totally closed after two weeks. (**H**) The self-repairing continued into the third week, with an apparent thickening over the freshly repaired hole.

We studied the self-repairing behavior by observing the regrowth of the holes that were deliberately drilled into the sample. The holes of ~5-20 nm in diameter (Fig. 2A) were created in a thin foil sample under examination in a Thermo-Fisher Themis Z300 operating at 300 kV using a converged electron beam focused onto thin regions of the specimen within the vacuum of the TEM. Firstly, self-repair was examined under ambient conditions. Once the holes were created, the samples were removed from the TEM column and stored at room temperature in air with an average humidity of 70%. Over the ensuing three weeks, the film was subsequently examined in the TEM. We observed a remarkable and gradual repairing of the damaged areas (Figs. 2B-2D). The low magnification TEM images presented in Figs. 2B-2D and the HRTEM images of hole 3 displayed in Figs. 2E-2H indicate a gradual self-repair of the holes in the weeks following the holes' formation. For example, hole 3 closed completely after two weeks (Fig. 2G). Once the self-repairing process had closed the hole, some further thickening at the regrown section was apparent, as can be seen by comparing Fig. 2H with Fig. 2G. This was also indicated by change in the image contrast from Fig. 2C to Fig. 2D in the original hole position. Interestingly, once the void of the hole has repaired, the newly regrown crystal maintains the original crystal structure and orientation.

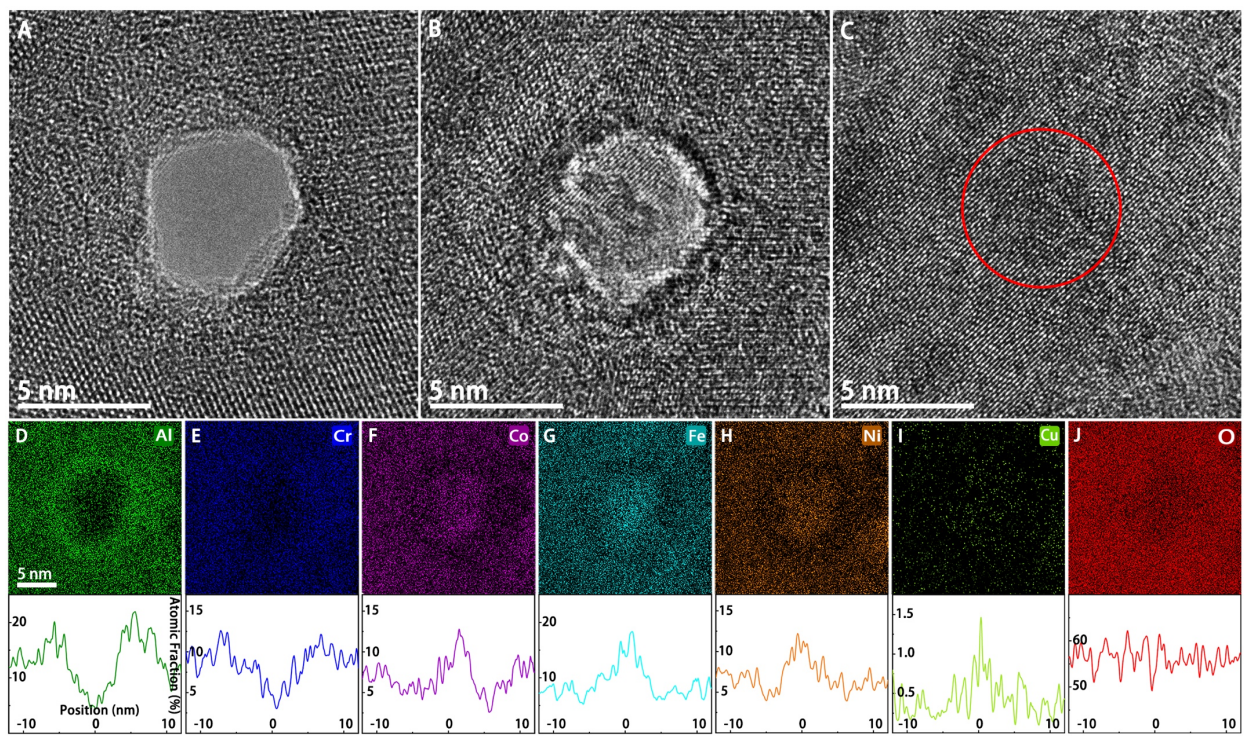

**Fig. 3. High-resolution TEM images of a hole 6 nm in diameter before and after self-repairing and the corresponding EDS results of the hole area**. (**A**) A high-resolution TEM image of a hole immediately after it was drilled with a converged 300 kV electron beam. The damage to the crystal lattice surrounding the hole is obvious and severe. This TEM specimen was removed from the TEM column and inserted into an oven heated to 300 °C in atmosphere for 30 min. After cooling to room temperature, the specimen was then re-inserted into the TEM and imaged as shown in (**B**). The hole had only partially regrown by this stage since the specimen was prematurely removed from the oven. Whilst under observation in the TEM, the self-repairing process continued and eventually resulted in a completely repaired specimen that appeared identical to its original, undamaged state (circled area in **C**). The crystal

orientation in the images in panels B and C is slightly different from that in panel A due to the re-attaching of the specimen to the TEM holder after heating in air. The EDS mapping and line scan results of the hole area after self-repairing are presented in (**D**) Al, (**E**) Cr, (**F**) Co, (**G**) Fe, (**H**) Ni, (**I**), Cu, and (**J**) O.

The regrowth kinetics demonstrated an exponential relationship with temperature. We observed that at 200 °C, the self-repairing occurred >100 times faster than the equivalent process at room temperature, and was >200 times faster at 300 °C in air. The 6 nm hole presented in Fig 3A grew back entirely within 30 min in air at 300 °C (Fig. 3B). Furthermore, we observed that energy forms other than thermal activation could also facilitate the regrowth—specifically, electron beam irradiation. When in the TEM under a 300 kV electron beam, only 35 min were required for a newly drilled 11 nm hole to be fully repaired. However, it is difficult to separate the respective contributions to the increased regrowth rate from, on the one hand, the excitation of particular atomic orbitals from irradiation by the high energy electrons and, on the other hand, the thermal heating from the electron beam. A recorded video of the self-repairing process observed in the TEM can be found in the Supplementary Materials.

To explore the detailed characteristics of the self-repairing process, energy-dispersive X-ray spectroscopy (EDS) was performed to analyze the compositional distributions of all elements in and around the original hole area after the regrowth was complete, as shown in Figs. 3D-3J. The EDS mapping of Al reveals a ring of higher Al concentration around the original hole (Fig. 3D, top). The EDS intensities of Co, Fe and Ni (Figs. 3F-3H, top row) were correspondingly weaker than that of their reference background intensities in the film. The trend of the Cr distribution (Fig. 3E) was broadly similar to that of Al as indicated by the EDS line scan across the hole (Fig. 3E, bottom). The correspondence between the higher concentrations of Al and Cr and the lower concentrations of Co, Fe and Ni around the edge of the hole is consistent with the enrichment observed in these latter species at the regrown part of the hole. Indeed, the newly regrown region of crystal that has filled the void of the hole contains an excess of Co, Fe and Ni and limited Al and Cr as displayed by the EDS mapping and line scan results (Fig. 3F-3H). Although the overall Cu content in the film is less than 0.5 at. %, its distribution (Fig. 3I) in the newly regrown area also follows the same trend as Co, Fe, and Ni. The change of elemental distribution implies that the regrowth was accompanied by a transition from the relatively random state that maximizes entropy in the as-received condition, to a chemically heterogeneous state that favors the release of stored enthalpy at moderate temperatures.

To fully understand the underlining mechanism of the self-repairing behavior, we employed density functional theory (DFT) calculations and material informatics methods to further explore the self-repairing process (Fig. 4). Our results suggest that high entropy oxides are intrinsically capable of self-repair because of the redistribution of entropy and enthalpy components of the Gibbs free energy (*14, 33, 34*):

$$G = H - TS \qquad (1)$$

Multi-cation high entropy oxides are typically prepared at high temperatures, where the entropy term dominates and the favored state of the distribution of the cations is highly disordered and is effectively random. At lower temperatures, where the entropy contribution is reduced (Eq. 1), it is the enthalpy term that governs the formation of the structure of the multi-cation oxide (Fig. 4B). The enthalpy, however, favors a very particular distribution of the cations. The cation sites in the spinel structure differentiate the tetrahedral A sites from the octahedral B sites. Generally, divalent cations, M(II), prefer the four-coordinated A sites while trivalent cations, M(III), favor the six-coordinated B sites. The swap between the M(II) and M(III) cations from their preferable sites would require additional energy (*35-38*). This means that upon cooling, the multi-cation oxide would be in a metastable state with cations randomly mixed between A and B sites. However, in the lattice without vacancies the relaxation into the stable ordered state is kinetically prohibited as swapping two cations requires a transition through a large energy barrier. However, in the presence of a gross defect such as the holes created reconstruction can be proceeded to form an ordered state.

In order to demonstrate that the release in enthalpy is the dominant driving force for the self-repairing in high-entropy oxides, we first calculated the formation enthalpies of different $AB_2O_4$ spinel structures for each pair among Al, Cr, Fe, Co, Ni, and Cu, as presented in Fig. 4C. It can be seen that Al and Cr are the most stable ones in the spinel and that there is a strong preference for Al and Cr to occupy the B sites in the spinel structure. We also calculated the formation enthalpies of the spinel structures in the form of $(M1_{0.5}M2_{0.5})(M3_{0.5}M4_{0.5}M5_{0.5}M6_{0.5})O_4$ where M1-M6 cover all possible combinations of Al, Cr, Fe, Co, Ni, and Cu, with M1 and M2 occupying the A sites and M3-M6 occupying the B sites. The results (Fig. 4D) indicate that Cu and Ni strongly favor their positioning at the A sites.

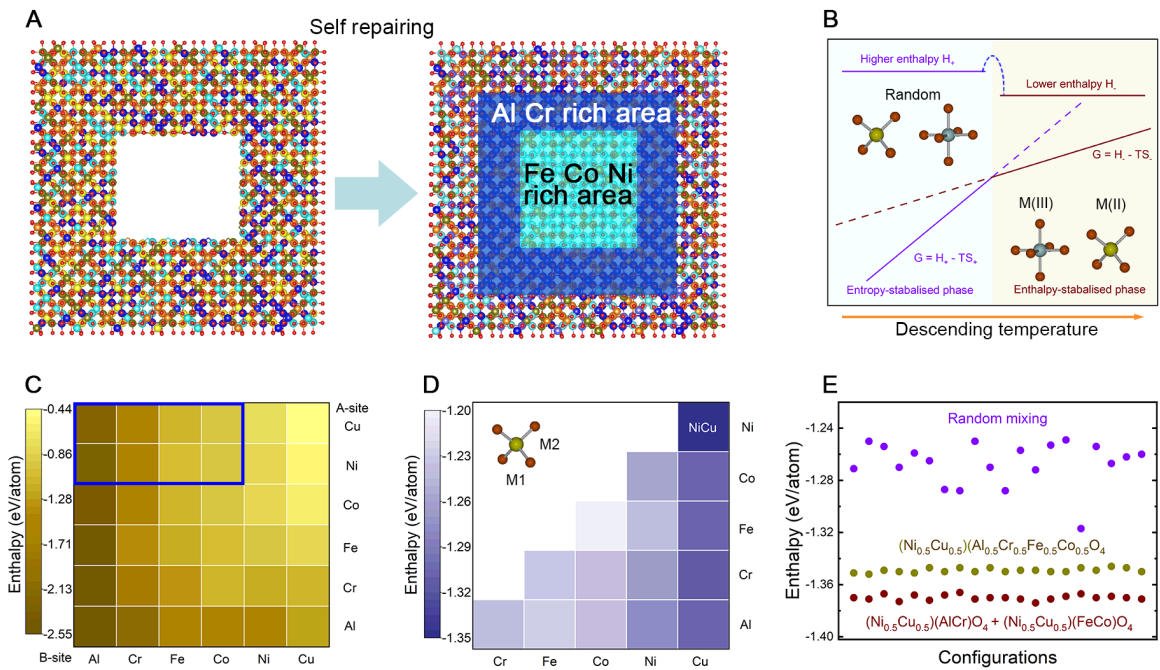

**Fig. 4. The underlying thermodynamics of the $(AlCoCrCu_{0.5}FeNi)_3O_4$ oxide film reconstruction**. (**A**) A schematic diagram for the reconstruction process of the high entropy oxide film involving the diffusion of elements from the region around the hole edge to the edge of the hole to form mainly Fe, Co and Ni based spinel oxide. This leaves the region around the edge enriched in Al and Cr in the spinel structure. Different colors represent different elemental species. (**B**) Phase diagram of multi-cation oxides involving a transition from high-temperature and high entropy-stabilized phase to the low-temperature enthalpy-stabilized phase. Both site ordering and phase separation are attributed to the enthalpy difference of the two phases. (**C**) The calculated formation enthalpies of different $AB_2O_4$ spinel structures for each pair among Al, Cr, Fe, Co, Ni, and Cu. (**D**) The formation enthalpy of equal-ratio six-cation spinel oxides in the form of $(M1_{0.5}M2_{0.5})(M3_{0.5}M4_{0.5}M5_{0.5}M6_{0.5})O_4$ where M1-M6 cover all possible combinations of Al, Cr, Fe, Co, Ni, and Cu, and M1 and M2 occupy the A sites and M3-M6 occupy the B sites. The diagram confirms that Ni and Cu are most favorable at tetrahedral sites at low temperature, corresponding to a composition highlighted (blue square) in **C** with a formula of $(Ni_{0.5}Cu_{0.5})(Fe_{0.5}Co_{0.5}Al_{0.5}Cr_{0.5})O_4$. (**E**)The calculated formation enthalpies of random mixing (purple dot), cation-ordering (olive, $(Ni_{0.5}Cu_{0.5})(Fe_{0.5}Co_{0.5}Al_{0.5}Cr_{0.5})O_4$) equal-ratio six-cation spinel oxides, and phase separated states (wine red, $(Ni_{0.5}Cu_{0.5})(AlCr)O_4$ and $(Ni_{0.5}Cu_{0.5})(FeCo)O_4$) spinel oxides. To show the trend of random assignments of cations, 20 configurations were calculated for each kind of structure.

The above calculations demonstrate that cations in the spinel oxide favor a specific site ordering to optimize the enthalpy of formation. This alone, however, is not enough to drive the self-repairing process, which requires simultaneous site ordering and phase separation to proceed. To prove that phase separation is indeed energetically favorable, we compared the energies of the partially site-ordered state of $(Ni_{0.5}Cu_{0.5})(Al_{0.5}Cr_{0.5}Fe_{0.5}Co_{0.5})O_4$ and phase separated states of $(Ni_{0.5}Cu_{0.5})(AlCr)O_4$ and $(Ni_{0.5}Cu_{0.5})(FeCo)O_4$ as shown in Fig. 4E. It is clear that such phase separation is energetically favorable, and the energy gain is mainly due to the release of atomic-level stresses and chemical heterogeneity associated with the disorder.

Based on our results, we propose the following description of the self-repairing process. Upon cooling from high-temperature crystal growth conditions, the high entropy oxides are frozen in a metastable state where all cations are arranged in near random distribution. In the absence of any major defects, such as a hole or a void, a transition to the energetically favorable ordered state is kinetically prohibited due to the large potential barrier for the swapping of cations between atomic sites and phase separation. However, the presence of a gross defect enables the ordering of the cations to proceed, providing a reaction pathway for phase separation and the process of self-repair. Specifically, if there is a rise in system free energy caused by the introduction of a hole, this effectively nucleates the growth of a new phase. In our case, the new phase was likely $(Ni_{0.5}Cu_{0.5})(FeCo)O_4$ in the hole, leaving the more stable configuration of $(Ni_{0.5}Cu_{0.5})(AlCr)O_4)$ at the edge of the hole. We believe that this is the general mechanism of the self-repairing process for a wide range of high entropy materials.

**References and Notes**

1. B. Cantor, I. T. H. Chang, P. Knight, A. J. B. Vincent. Microstructural development in equiatomic multicomponent alloys. *Mater. Sci. Eng. A* **375–377**, 213-218 (2004).

2. J. W. Yeh, S. K. Chen, S. J. Lin, S. J. Y. Gan, T. S. Chin, T. T. Shun, C. H. Tsau, S. Y. Chang. Nanostructured high−entropy alloys with multiple principal elements: novel alloy design concepts and outcomes. *Adv. Eng. Mater.* **6,** 299-303 (2004).

3. Y. G. Yao, Z. N. Huang, P. F. Xie, S. D. Lacey, R. J. Jacob. Carbothermal shock synthesis of high-entropy-alloy nanoparticles. *Science* **359**, 1489–1494 (2018).

4. M. Naeem, H. Y. He, F. Zhang, H. L. Huang, S. Harjo et al. Cooperative deformation in high-entropy alloys at ultralow temperatures. *Sci. Adv*. **6**: eaax4002 (2020).

5. B. Gludovatz, A. Hohenwarter, D. Catoor, E. H. Chang, E. P. George, R. O. Ritchie. A fracture-resistant high-entropy alloy for cryogenic applications. *Science* **345**, 1153-1158 2014).

6. C. C. Tung, J. W. Yeh, E. T. Shun, S. K. Chen, Y. S. Huang, C. C. Chen. On the elemental effect of AlCoCrCuFeNi high-entropy alloy system. *Mater. Lett.* **61**, 1-5 (2007).

7. Y. Z. Shi, B.Yang, P. K. Liaw. Corrosion-Resistant High-Entropy Alloys: A Review. *Metals* **7**, 43 (2017).

8. S. Xia, Z. Wang, T. F. Yang, Y. Zhang. Irradiation behavior in high entropy alloys. *J. Iron Steel Res. Int.* **22**, 879-84 (2015).

9. S. Xia, X. Yang, T. F. Yang, S. Liu, Y. Zhang. Irradiation resistance in $Al_x$CoCrFeNi high entropy alloys. *Jom.* **67**, 2340-4 (2015).

10. S. Xia, M. C. Gao, T. Yang, P. K. Liaw, Y. Zhang. Phase stability and microstructures of high entropy alloys ion irradiated to high doses. *J. Nucl. Mater.* **480**, 100-8 (2016).

11. L. Yang, H. Ge, J. Zhang, T. Xiong, Q. Jin, Y. Zhou, et al. High He-ion irradiation resistance of CrMnFeCoNi high-entropy alloy revealed by comparison study with Ni and 304SS. *J. Mater. Sci. Technol.* **35**, 300-5 (2019).

12. C. Oses, C. Toher, S. Curtarolo. High-entropy ceramics. *Nature* **5**, 295-309 (2020).

13. A. Sarkar, Q. Wang, A. Schiele, MR. Chellali, S.S. Bhattacharya, D. Wang, et al. High‒entropy oxides: fundamental aspects and electrochemical properties. *Adv. Mater.* **31**, 1806236 (2019).

14. J. Gild, Y. Zhang, T. Harrington, S. Jiang, T. Hu, M. C. Quinn,W. M. Mellor, N. Zhou, K. Vecchio, J. Luo. High-entropy metal diborides: a new class of high-entropy materials and a new type of ultrahigh temperature ceramics. *Sci Rep.* **6**, 1-10 (2016).

15. B. B. Jiang, Y. Yu, J. Cui, X.X. Liu, L. Xie et al. High-entropy-stabilized chalcogenides with high thermoelectric performance. *Science* **371**, 830-834 (2021).

16. E. Castle, T. Csanádi, S. Grasso, J. Dusza, M. Reece. Processing and properties of high-entropy ultra-high temperature carbides. *Sci Rep.* **8**, 1-12 (2018).

17. R. Z. Zhang, F. Gucci, H. Zhu, K. Chen, Reece MJ. Data-driven design of ecofriendly thermoelectric high-entropy sulfides. *Inorg. Chem.* **57**, 13027-33 (2018).

18. T. Jin, X. Sang, R. R. Unocic, R. T. Kinch, X. Liu, J. Hu, H. Liu, S. Dai. Mechanochemical-assisted synthesis of high-entropy metal nitride via a soft urea strategy. *Adv. Mater.* **30**, 1707512 (2018).

19. N. A. Khan, B. Akhavan, C. F. Zhou, H. R. Zhou, L. Chang et al. High entropy nitride (HEN) thin films of AlCoCrCu0. 5FeNi deposited by reactive magnetron sputtering. *Surf. Coat. Technol*. **402**, 1263273 (2020).

20. R. Djenadic, A. Sarkar, O. Clemens, C. Loho, M. Botros, V.S.K. Chakravadhanula, C. Kuebel, S.S. Bhattacharya, A.S. Gandhif, H. Hahn. Multicomponent equiatomic rare earth oxides. *Mater. Res. Lett.* **5**, 102-9 (2017).

21. A. Sarkar, C. Loho, L. Velasco, T. Thomas, S. S. Bhattacharya, H. Hahn, R. Djenadic. Multicomponent equiatomic rare earth oxides with a narrow band gap and associated praseodymium multivalency. *Dalton Trans*. **46**, 12167-76 (2017).

22. S. Jiang, T. Hu, J. Gild, N. Zhou, J. Nie, M. Qin, T. Harrington, K. Vecchio, J. Luo. A new class of high-entropy perovskite oxides. *Scr. Mater.* **142**, 116-20 (2018).

23. A. Sarkar, R. Djenadic, D. Wang, C. Hein, R. Kautenburger, O. Clemens, H. Hahn. Rare earth and transition metal based entropy stabilised perovskite type oxides. *J. Eur. Ceram. Soc*. **38**, 2318-27 (2018).

24. J. Dąbrowa, M. Stygar, A. Mikuła, A. Knapik, and K. Mroczka, W. Tejchman, M. Danielewski, M. Martin. Synthesis and microstructure of the (Co, Cr, Fe, Mn, Ni)$_3$O$_4$ high entropy oxide characterized by spinel structure. *Mater. Lett.* **216**, 32-6 (2018).

25. A. Mao, F. Quan, H. Xiang, Z. Zhang, K. Kuramoto, A. Xia. Facile synthesis and ferrimagnetic property of spinel (CoCrFeMnNi)$_3$O$_4$ high-entropy oxide nanocrystalline powder. *J. Mol. Struct.* **1194**, 11-8 (2019).

26. D. Wang, S. Jiang, C. Duan, J. Mao, Y. Dong, K. Don, Z. Wang, S. Luo, Y. Liu, X. Qi. Spinel-structured high entropy oxide (FeCoNiCrMn)$_3$O$_4$ as anode towards superior lithium storage performance. *J. Alloy. Compd.* **844**, 156158 (2020).

27. N. A. Khan, B. Akhavan, Z. Zheng, H. Liu, C. Zhou, H. Zhou, L. Chang, Y. Wang, Y. Liu, L. Sun, M. M. Bilek, Z. Liu. Nanostructured AlCoCrCu$_{0.5}$FeNi high entropy oxide (HEO) thin films fabricated using reactive magnetron sputtering. *Appl. Surf. Sci.* **553**, 149491 (2021).

28. N. A. Khan, B. Akhavan, H. Zhou, L. Chang, Y. Wang, L. Sun, M.M. Bilek, Z. Liu. High entropy alloy thin films of AlCoCrCu$_{0.5}$FeNi with controlled microstructure. *Appl. Surf. Sci.* **495**, 143560 (2019).

29. N. A. Khan, B. Akhavan, C. Zhou, H. Zhou, L. Chang, Y. Wang, Y. Liu, L. Fu, M.M. Bilek, Z. Liu. RF magnetron sputtered AlCoCrCu$_{0.5}$FeNi high entropy alloy (HEA) thin films with tuned microstructure and chemical composition. *J. Alloy. Compd.* **836**, 155348 (2020).

30. Y. S. Huang, L. Chen, H. W. Lui, M. H. Cai, J. W. Yeh. Microstructure, hardness, resistivity and thermal stability of sputtered oxide films of AlCoCrCu$_{0.5}$NiFe high-entropy alloy. *Mater. Sci. Eng., A*. **457**, 77-83 (2007).

31. T. K. Chen. M. S. Wong. Structure and properties of reactively-sputtered Al$_x$CoCrCuFeNi oxide films. *Thin Solid Films*. **516**, 141-6 (2007).

32. T. K. Chen. M. S. Wong. Thermal stability of hard transparent Al$_x$CoCrCuFeNi oxide thin films. *Surf Coat Tech.* **203**, 495-500 (2008).

33. C. M. Rost, E. Sachet, T. Borman, A. Moballegh, E.C. Dickey, D. Hou, J.L. Jones, S. Curtarolo, J.-P. Maria. Entropy-stabilized oxides. *Nat. Commun.* **6**, 1-8 (2015).

34. S. J. McCormack, A. Navrotsky. Thermodynamics of high entropy oxides. *Acta Mater.* **202**, 1–21 (2021).

35. C. Biagioni, M. Pasero. The systematics of the spinel-type minerals: An overview. *Am. Miner*. **99**, 1254-64 (2014).

36. K. E. Sickafus, JM. Wills, NW. Grimes. Structure of spinel. *J. Am. Ceram. Soc.* **82**, 3279-92 (1999).

37. Q. Zhao, Z. Yan, C. Chen, J. Chen. Spinels: controlled preparation, oxygen reduction/evolution reaction application, and beyond. *Chem. Rev.* **117**, 10121-211 (2017).

38. Z. Szotek, W. M. Temmerman, D. Ködderitzsch, A. Svane, L. Petit, and H. Winter. Electronic structures of normal and inverse spinel ferrites from first principles. *Phys. Rev. B*. **74**, 174431 (2006).

39. J. P. Perdew, K. Burke, M Ernzerhof. Generalized gradient approximation made simple. *Phys. Rev. Lett.* **77**, 3865 (1996).

40. G. Kresse, J. Furthmüller. Efficient iterative schemes for ab initio total-energy calculations using a plane-wave basis set. *Phys. Rev. B*. **54**, 11169 (1996).

41. G. Kresse, D. Joubert. From ultrasoft pseudopotentials to the projector augmented-wave method. *Phys. Rev. B*. **59**, 1758 (1999).

42. P. E. Blöchl. Projector augmented-wave method. *Phys. Rev. B*. **50**, 17953 (1994).

43. A. Jain, S. P. Ong, G. Hautier, W. Chen, W.D. Richards, S. Dacek, S. Cholia, D. Gunter, D. Skinner, G. Ceder, K.A. Persson. The Materials Project: A materials genome approach to accelerating materials innovation. *APL Mater.* **1**, 011002 (2013).

**Acknowledgments:** The authors wish to thank Dr Ian Falconer and Mr Julian Whichello of School of Physics at the University of Sydney for their support in material deposition by radio frequency (RF) magnetron sputtering. The authors acknowledge the facilities, as well as the scientific and technical staff support of Sydney Analytical and Sydney Microscopy & Microanalysis (SMM) at the University of Sydney (core research facilities). SMM is a node of Microscopy Australia. The authors acknowledge particularly the TEM support from Drs. Hongwei Liu and Magnus Garbrecht of SMM.

**Funding:** ZL acknowledges the funding support from the Australian Research Council (ARC Discovery Projects, DP210103539, DP180102976 and DP130104231).

**Author contributions:** ZL designed the project and directed the research. PH, KSN and LS caried out the DFT calculations. JQ and ZZ performed the TEM characterizations. NAK synthesized the high entropy oxide thin films. MMB, BA and KT guided the film deposition.

CZ assisted the cross-sectional TEM specimen preparation. All the authors participated in the discussions of the research work that contributed to the conclusions of the manuscript.

**Competing interests:** The authors declare that they have no competing interests.

**Data and materials availability:** All data are available in the main text or the supplementary materials.

# Supplementary Materials for

## Self-repairing high entropy oxides

Zongwen Liu*†, Pengru Huang†, Lixian Sun*, Yanping Liu, Jiangtao Qu, Julie Cairney, Zhong Zheng, Zhiming M. Wang, Naveed A. Khan, Zhiping Lai, Li Fu, Bing Teng, Cuifeng Zhou, Hong Zhao, Fen Xu, Pan Xiong, Junwu Zhu, Peng Yuan, Kosta Tsoutas, Behnam Akhavan, Marcela M. Bilek, Simon P. Ringer, Kostya S. Novoselov*

*Correspondence to: zongwen.liu@sydney.edu.au; sunlx@guet.edu.cn; kostya@nus.edu.sg
†These authors contributed equally to this work.

**Materials and Methods**

High entropy thin film fabrication

The (AlCoCrCu$_{0.5}$FeNi)$_3$O$_4$ high entropy oxide film was prepared by RF magnetron sputtering. The target of AlCoCrCu0.5FeNi was prepared by the arc melting and casting method. The sputtered (AlCoCrCu$_{0.5}$FeNi)$_3$O$_4$ was deposited on a Si (100) single crystal wafer. The silicon wafer was cleaned with water and acetone prior to deposition. The power on the AlCoCrCu$_{0.5}$FeNi target was 200 W. The deposition was performed at a total pressure of 10$^{-3}$ Torr in a mixed Ar and O$_2$ atmosphere. The ratio between the oxygen partial pressure and the total working gas pressure (pO$_2$% = pO$_2$/(pO$_2$ + pAr)) was 30 %. The substrate temperature was kept at 300 °C. The subsequent annealing treatment was carried out on a vacuum furnace at 900 °C.

DFT Calculations

Spin-polarized density functional theory calculations were performed to obtain the formation enthalpies of spinel structures using the Perdew, Burke, and Ernzerhof (PBE) functional as implemented in the Vienna Ab Initio Simulation Package (VASP) (*39-41*). The interaction between the valence electrons and ionic cores was described within the projector augmented (PAW) approach with a plane-wave energy cutoff of 500 eV (*42*). The Brillouin zone was sampled using a (6×6×6) Monkhorst-Pack grid for the cubic unit cell. In the structural energy minimization, the atomic coordinates were allowed to relax until the forces on all the atoms were less than 0.01 eV/Å. The computational work for this article was fully performed on resources at the National Supercomputing Centre, Singapore (https://www.nscc.sg).

**Supplementary Text**

The formation enthalpy of spinel oxides

As revealed by the structural characterization, the high entropy oxide possesses a spinel structure, and we adopted this structure as the host crystal for our calculations. The normal spinel structure with the formula of $AB_2O_4$ is composed of tetrahedrally coordinated A-site cations with four nearest oxygens and B-site cations octahedrally bonded with six oxygens. As show in Fig. S1, the unit cell of the spinel contains 8 A-cation sites, 16 B-cation sites and 32 oxygen sites. The formation enthalpies of the structures were calculated according to

$$E_f = E(spinel) - n_A E_A - n_B E_B - n_O E_O \qquad (1)$$

where E(*spinel*) is the energy of the spinel oxides, including the binary, random mixing, partially ordering, or phase-separation multi-cation structures discussed in the main text, while $E_A$ and $E_B$ are the energies of A- and B-site cations in their solid element forms, respectively. $E_O$ is the energy of an oxygen atom in the gas form. $n_A$, $n_B$, and $n_O$ are the numbers of the A cations, B cations, and oxygens included in the structure. To simulate the totally or partially cation-mixing high entropy multi-cation oxides, different elements were randomly assigned in specific cation sites.

To support the formation enthalpies calculated in the main text, we collected the corresponding data from the Materials Genome Initiative database of Material projects (*43*), as shown in Fig. S2. It is evident, that our data of Fig. 4c is qualitatively consistent with that in the database, proving that Al and Cr are most stable in the spinel crystal and there is a strong preference for Al and Cr to occupy the B sites in the spinel structure.

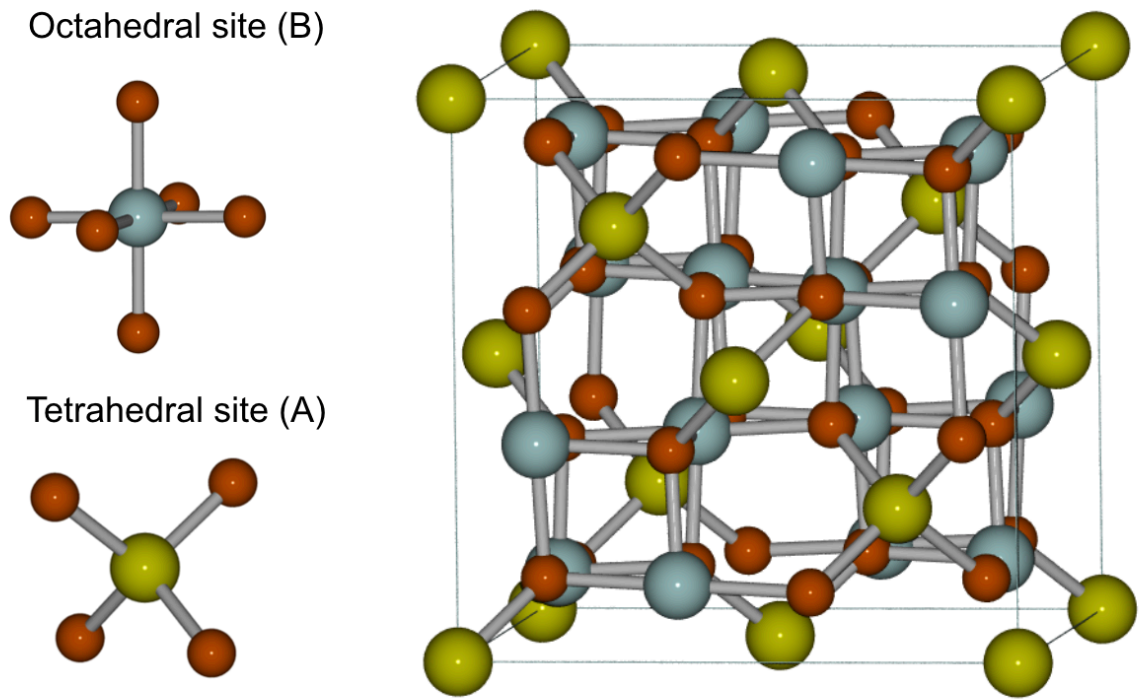

**Fig. S1.**
**The binary cation-site feature of the spinel crystal structure**. The cubic unit cell of the spinel structure has two different cation sites, the tetrahedral A sites and the octahedral B sites. Generally, some elements favor the four-coordinated A sites while the others prefer the six-coordinated B sites. Swapping of the cations from their favorable sites would result in an enthalpy increase, making more energy stored in the multi-cation oxide of spinel.

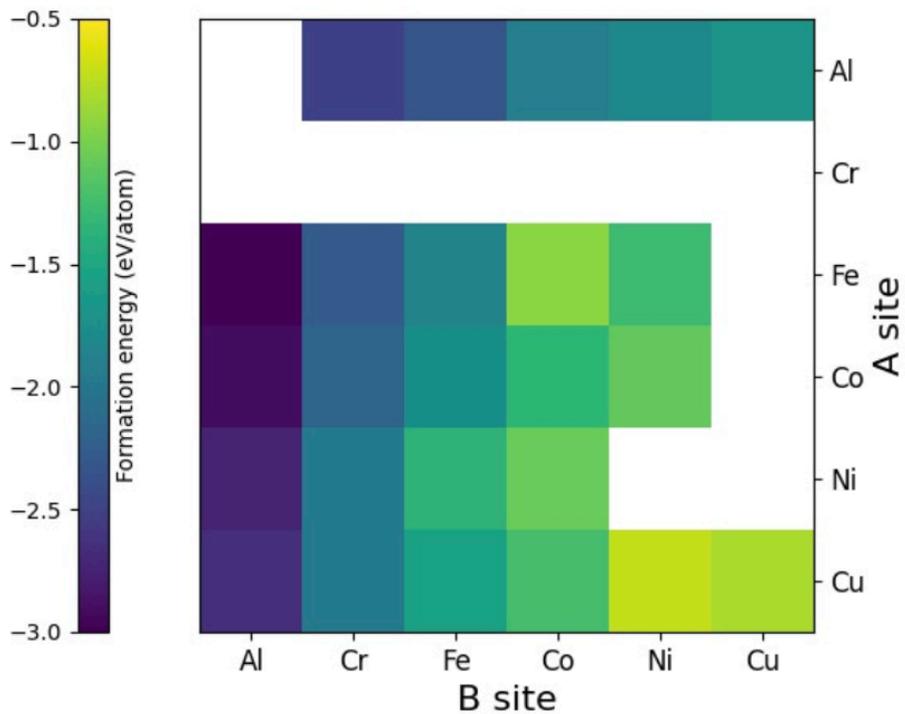

**Fig. S2.**

**The formation energy per atom collected from the Materials Genome Initiative database of Material projects**. The collect formation energies for each pair of cations, among Al, Cr, Fe, Co, Ni, and Cu, show that Al and Cr are most stable in the spinel structure and favor the B site. The blank data are due to the lack of information in the database.

**Movie S1.**

**A video showing the self-repairing process of AlCoCrCu$_{0.5}$FeNi)$_3$O$_4$ high entropy oxide film recorded in a TEM.** First, a hole of 11 nm in diameter was created by a converged 300kV electron beam, then the electron beam was spread out and the crystal growth around the edge of hole could be seen vividly. While it took 35 minutes for the hole to fully grow back, the video was compressed to 53 seconds.